\documentclass[aps,prb,twocolumn,showpacs,amsmath,amssymb,amsthm,superscriptaddress,groupedaddress]{revtex4}
\usepackage{graphics}
\usepackage{epsfig}
\usepackage[colorlinks=true,citecolor=blue,linkcolor=red,linktocpage=true,pagebackref=false]{hyperref}
\usepackage{soul} 
\usepackage[usenames, dvipsnames]{color}
\usepackage{braket}

\begin{document}
\title{
Single-particle shot noise at non-zero temperature
}
\author{Michael Moskalets}
\email{michael.moskalets@gmail.com}
\affiliation{Department of Metal and Semiconductor Physics, NTU ``Kharkiv Polytechnic Institute", 61002 Kharkiv, Ukraine}

\date\today
\begin{abstract}
The state of a single particle injected onto the surface of the Fermi sea is a pure state if the temperature is zero and is a mixed state if the temperature is finite. 
Moreover, the state of an injected particle is orthogonal to the state of the  Fermi sea at zero temperature, while it is not orthogonal at non-zero temperature. 
These changes in the quantum state of the injected particles can be detected using the temperature dependence of the shot noise that is generated when the particles one by one pass through a semitransparent quantum point contact. 
Namely, the shot noise produced by the mixed state is suppressed in comparison with the noise of the pure state.  
In addition, the correlations between the injected particles and the underlying Fermi sea, present at non-zero temperature, do enhance the shot noise. 
Furthermore, antibunching of injected particles with possible thermal excitations coming from another input channel of a quantum point contact  does suppress shot noise. 
Here I analyze in detail these three effects, which are responsible for the temperature dependence of the shot noise, and discuss how to distinguish them experimentally.  
\end{abstract}
\maketitle

\section{Introduction}

Recent implementation of on-demand coherent single-electron sources \cite{Feve:2007jx,Dubois:2013ul} and subsequent demonstration of the quantum tomography of a single-electron wave function \cite{Jullien:2014ii} and of the controllable two-particle interference \cite{{Dubois:2013ul},{Bocquillon:2013dp},{Marguerite:2016ur},{Glattli:2016wl}} mark reaching the fundamental limit of the solid-state coherent electronics. 
This is the limit when the quantum properties of the smallest possible individual  carriers manifest themselves directly in transport phenomena. 

Here I address the question of how the state of a particle generated by a single-electron source is affected by the Fermi sea. 
To be more specific, I am interested in the physical phenomena accompanying injection of single particles on top of the Fermi sea at non-zero temperature  and how they manifest themselves in transport. 

This question was already addressed experimentally. 
Thus, the temperature dependence of the shot noise of the single particles (electrons and holes) generated by a quantum capacitor \cite{Buttiker:1993wh,{Gabelli:2006eg}} was reported in Ref.~\onlinecite{Bocquillon:2012if}. 
It was demonstrated experimentally that the shot noise gets suppressed with increasing temperature. 
The same effect was predicted \cite{{Dubois:2013fs}} and observed \cite{{Glattli:2016tr},Glattli:2016wl} when elementary excitations, levitons, were generated with the help of Lorentzian voltage pulses \cite{Levitov:1996ie,Ivanov:1997wz,{Keeling:2006hq}}.

The temperature-induced shot noise suppression was explained as a result of quantum-statistical, fermionic antibunching of injected particles and thermally excited (quasi-)electrons and holes existing in the Fermi sea at non-zero temperature. \cite{Bocquillon:2012if,{Bocquillon:2013fp}}

Another explanation of this effect was put forward in Ref.~\onlinecite{Moskalets:2015ub}, where the shot noise suppression was related to the fact that the single-particle state injected at non-zero temperature is a mixed state, which is generally less noisy compared to a pure state.  

Here I argue that in addition to antibunching with thermal excitations and thermal mixing of the single-particle state, which both suppress the shot noise,  there is additional effect, which enhances the shot noise. 

This latter effect is related to the fact that at non-zero temperature the state of an  injected particle is not orthogonal to the states of the Fermi sea.
Although  the fermionic quantum-statistical repulsion enforces injected particles and particles of the Fermi sea to be in different states at each time, it does not prevent them from occupying the same state but at different times. 
Indeed, being in the mixed state, a particle occupies different components of this mixed state at different times. 
Accordingly, two particles, each being in its own mixed state, can have coincident (or, better to say, not orthogonal) components, which they occupy at different times.

As I show below, the contribution to the shot noise arising due to correlations between injected particles and the thermal excitations of the Fermi sea is opposite to that of the antibunching effect.
I propose experimental setups allowing to address each effect separately. 
 
The paper is organized as follows. 
In Sec.~\ref{sec2} the temperature dependence of the shot noise is analyzed on the basis of the general relationship between the shot noise and the quantum state of scattered particles. 
I identify three physical mechanisms that influence this dependence, and in Sec.~\ref{sec2d} discuss how to separate their contributions experimentally.
As an example, in Sec.~\ref{sec3} the temperature dependence of the shot noise of levitons is analyzed in detail. 
An account of results obtained in this work and their short discussion are given in Sec.~\ref{concl}. 

\section{Shot noise}
\label{sec2}

\subsection{Setup}

\begin{figure}[t]
\includegraphics[width=80mm, angle=0]{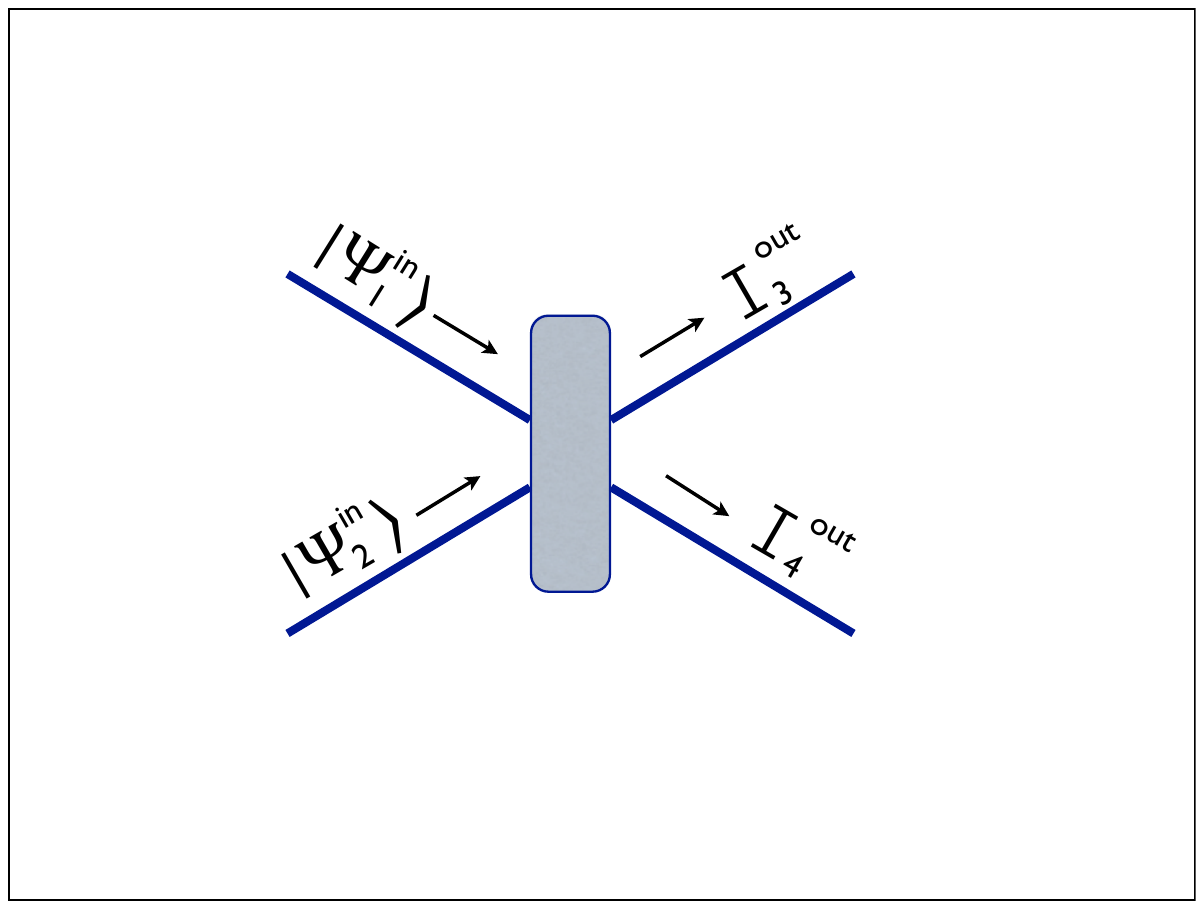}
\caption{A mesoscopic electron collider circuit with two input and two output chiral electron waveguides (shown as blue straight lines)  coupled via an electron wave splitter (shown as a rounded rectangle). Two incoming states $ \ket{ \Psi_{ \alpha}^{in}}$, $ \alpha = 1,2$, are scattered on the wave splitter. As a result the two outgoing currents $I_{ \beta}^{out}$, $ \beta = 3,4$ are generated. }
\label{fig1}
\end{figure}

I consider a mesoscopic collider \cite{{Buttiker:1992vr},Liu:1998wr,{Henny:1999tb}}, an electron circuit consisting of two input and two output single-channel chiral electron waveguides and an electron wave splitter, see Fig.~\ref{fig1}. 
These wave-guides can be, for instance, the chiral edge states of conductors in the integer quantum Hall effect regime. \cite{Klitzing:1980kwa,Halperin:1982tb,Buttiker:1988vk}
The wave splitter can be a quantum point contact. \cite{VanWees:1988vf,{Wharam:1988vi}}

The input waveguides are originated from their corresponding metallic contacts kept at the same or different temperatures, $ \theta_{ \alpha}$, $ \alpha = 1, 2$, and at the same chemical potential, $ \mu_{ \alpha} = \mu $. 
Thus, the (noninteracting) electron systems of input waveguides are in equilibrium and they are described by the respective Fermi distribution functions $f_{ \alpha}\left(  \epsilon \right) = \left(  1 + \exp\left[ \frac{ \epsilon }{ k_{B} \theta_{ \alpha}  } \right] \right)^{-1}$ with $k_{B}$ being the Boltzmann constant and $ \epsilon = E - \mu$ being the energy $E$ measured from the Fermi energy $ \mu$.    
At zero temperature all states with $\epsilon< 0$ are fully occupied and all states with $\epsilon> 0$ are empty.   
While, at non-zero temperature the states with $\epsilon> 0$ are partially occupied and states with $\epsilon< 0$ are partially emptied. 
It is convenient to treat the Fermi sea at zero temperature as (quasi-particle) vacuum. 
Then, the Fermi sea at non-zero temperature can be considered as vacuum with excitations. 
The thermal excitations with $\epsilon> 0$ are quasi-electrons, while the empty states with $\epsilon< 0$ are holes. 

In addition, there is a single-electron source \cite{Feve:2007jx,Dubois:2013ul}, which injects particles at a  rate of one particle per period ${\cal T}$  into the input $ \alpha =1$. 
Therefore, the quantum state $ \ket{ \Psi^{in}_{1}}$, impinging the wave splitter through the channel $ \alpha =1$, (see, Fig.~\ref{fig1}) describes  thermal excitations, quasi-electrons and holes, at the temperature $ \theta_{1}$ and particles injected by a single-electron source. 
While the quantum state $ \ket{ \Psi^{in}_{2}}$, impinging the wave splitter through the channel $ \alpha =2$, describes thermal excitations at the temperature $ \theta_{2}$ only. 

The incoming particles are scattered by the wave splitter into the outputs, where they generate electrical currents, $I_{ \beta}^{out}$, $ \beta = 3, 4$, see Fig.~\ref{fig1}. 
The measurement of these currents and their fluctuations provides information on the quantum states of the incoming particles.

\subsection{An electrical noise and the quantum state of electrons}

I am interested in the long-time correlations that arise between fluctuating outgoing currents, $I_{3}^{out}(t_{1})$ and $I_{4}^{out}(t_{2})$. 
These correlations are characterized via the current-current cross-correlation function integrated over the time difference $t_{1} - t_{2}$ and  averaged over the mean time, $\left(  t_{1} + t_{2} \right)/2$. \cite{Blanter:2000wi} 
I denote this electrical current correlation function as ${\cal P}_{34}$.

When individual particles, electrons or holes, are scattered on a wave splitter, the outgoing currents fluctuate. 
Since an elementary indivisible particle cannot be scattered to both outputs at once, the fluctuations of outgoing currents are anti-correlated and the correlation function ${\cal P}_{34}$ is negative. 
This source of current fluctuations is called \emph{shot noise}.\cite{Schottky1918,Li:1990es,Liefrink:1991es,Reznikov:1995us,Saminadayar:1997dw,dePicciotto:1997dk,Reznikov:1998kn,Blanter:2000wi}  
For brevity, I will use the same name, the shot noise, also for current fluctuations and for the current-current correlation function ${\cal P}_{34}$. 

When the incoming particles are in a pure quantum state, the shot noise is not sensitive to the details of the wave function. 
In contrast, when the incoming particles are in a mixed quantum state, this is no longer so: The shot noise gets suppressed and the value of the current correlation function depends on the properties of the mixed state. \cite{Moskalets:2015ub}

Another factors that affect the shot noise are correlations of two types: (i)  correlations that arise when an electron is injected into the waveguide with other particles, and (ii) correlations that arise when particles from different waveguides collide on a wave splitter. 
At non-zero temperature, we inevitably encounter all the aforementioned factors.

To analyze the shot noise at non-zero temperature, it is convenient to express the shot noise  ${\cal P}_{34}$ in terms of the first-order fermionic  correlation function, $ G ^{(1)}_{ \alpha}$ \cite{{Grenier:2011js},{Grenier:2011dv},{Haack:2013ch},Moskalets:2013dl} for incoming particles in the input channels $ \alpha = 1, 2$. 
The corresponding equation in the case of equal temperatures, $ \theta_{1} = \theta_{2}$, is obtained in Ref.~\onlinecite{Moskalets:2015ub} and it can  straightforwardly be generalized  to the case of unequal temperatures,
\begin{eqnarray}
\frac{  {\cal P}^{}_{34} }{  {\cal P}_{0} } &=& - v_{ \mu}^{2} \iint\limits dt_{1} dt_{2}\left |   G ^{(1)}_{1}\left(t_{1};t_{2}\right) - G ^{(1)}_{ 2}\left(t_{1};t_{2}\right)  \right |^{2} .
\label{01} 
\end{eqnarray}
\ \\ \noindent
Here ${\cal P}_{0} = e^{2}T\left(  1-T \right)/ {\cal T}$ with $T$ being the transmission probability of a wave splitter.  
The factor $v_{ \mu}^{2}$, the square of an electron velocity, comes from the normalization of fermionic correlation functions.  

The correlation functions $ G ^{(1)}_{ \alpha}$ for the incoming state $ \ket{ \Psi^{in}_{ \alpha}}$  can be represented as follows, 

\begin{eqnarray}
G ^{(1)}_{1} &=& G ^{(1)}_{p, \theta_{1}} + G ^{(1)}_{th,\theta_{1}} + G ^{(1)}_{0} , 
\nonumber \\
\label{02} \\
G ^{(1)}_{2} &=& G ^{(1)}_{th,\theta_{2}} + G ^{(1)}_{0} , 
\nonumber 
\end{eqnarray}
\noindent \\
where $G ^{(1)}_{p, \theta_{1}}$ is the first-order fermionic correlation function of particles injected by the single-electron source into the waveguide at the temperature $ \theta_{1}$, \cite{Moskalets:2015ub,Moskalets:2017vk}
$G ^{(1)}_{th, \theta_{ \alpha}}$ is the  first-order fermionic correlation function of thermal excitations at the temperature $ \theta_{\alpha}$, 
and, finally, $ G ^{(1)}_{0}$ is the  first-order fermionic correlation function of  vacuum, that is of the Fermi sea at zero temperature and chemical potential $ \mu$. 

Apparently, the vacuum does not contribute to the shot noise, Eq.~(\ref{01}).

\subsection{Excess shot noise}

When the single-particle source is turned off and $ G ^{(1)}_{p, \theta_{1}} =0$, the shot noise, however, is generally not zero. 
Its source, at $ \theta_{1} \ne \theta_{2}$, is the difference of fluxes of thermal excitations impinging the wave splitter from different input channels. 
Such a shot noise, caused by the thermal excitations, should not be confused with the thermal noise \cite{Johnson1928,Nyquist1928,Blanter:2000wi}, which does not contribute to the current cross-correlation function ${\cal P}^{}_{34}$ in the setup shown in Fig.~\ref{fig1}. 
In any case, the shot noise caused by thermal excitations is not of a prime interest here. 

The focus of this work is on the contribution caused by the particles generated by the source. 
This contribution is defined as the difference of ${\cal P}^{}_{34}$ with the particle source being switched on and off and is called {\it the excess shot noise},
\begin{eqnarray}
 {\cal P}^{ex}_{34} &=& \left(   {\cal P}^{}_{34}\right)_{on} - \left(   {\cal P}^{}_{34} \right)_{off} .
\label{03}
\end{eqnarray}
\noindent \\ 
After substituting Eq.~(\ref{02}) into Eq.~(\ref{01}) and then into Eq.~(\ref{03}),  the excess noise is represented as the sum of three contributions, 
\begin{subequations}
\label{04}
\begin{eqnarray}
{\cal P}^{ex}_{34} &=& {\cal P}^{p}_{34}  +  {\cal P}^{corr}_{34} +  {\cal P}^{ab}_{34}.
\label{04a} 
\end{eqnarray}
\noindent \\
Here 
\begin{eqnarray}
\frac{  {\cal P}^{p}_{34} }{  {\cal P}_{0} } &=& - v_{ \mu}^{2} \iint\limits dt_{1} dt_{2}\left |   G ^{(1)}_{p, \theta_{1}}\left(t_{1};t_{2}\right)  \right |^{2} ,
\label{04b} 
\end{eqnarray}
\noindent \\
is the main part of the shot noise, caused solely by the scattering of particles generated by the source on the wave splitter.

At zero temperature and in the single-particle emission regime the left hand side is just $-1$. 
This becomes clear if to remind that in this case $G ^{(1)}_{p, \theta_{1}=0}\left(t_{1};t_{2}\right) =  v_{ \mu}^{-1} \Psi^{*}\left(  t_{1} \right)\Psi^{}\left(  t_{2} \right)$, where $ \Psi(t)$ is the wave function of an injected electron (normalized to one), see, e.g., Refs.~\onlinecite{Grenier:2011js,Grenier:2013gg,Moskalets:2015ub}.  

At non-zero temperature, the state of an injected electron is a mixed state. 
Since the components of mixed state are occupied with probability less then one, the absolute value of the integral on the left hand side is reduced. \cite{Moskalets:2015ub} 
That is, with increasing temperature the injected particles become less noisy.

The two other terms, the correlation correction, 

\begin{eqnarray}
\frac{  {\cal P}^{corr}_{34} }{  {\cal P}_{0} } &=&  -2J_{1},
\label{04c} 
\end{eqnarray}
\noindent \\
and the antibunching correction, 

\begin{eqnarray}
\frac{  {\cal P}^{ab}_{34} }{  {\cal P}_{0} } &=& 2J_{2} ,
\label{04d} 
\end{eqnarray} 
\noindent \\
are defined by the following quantity, 

\begin{eqnarray}
J_{ \alpha} &=&  v_{ \mu}^{2} \iint\limits dt_{1} dt_{2} G ^{(1)}_{p, \theta_{1}}\left(t_{1};t_{2}\right) \left[ G ^{(1)}_{th, \theta_{\alpha}}\left(t_{1};t_{2}\right) \right]^{*} ,  
\label{04e} 
\end{eqnarray}
\end{subequations}
\noindent \\
which characterizes {\it the second-order coherence} between injected particles and thermal excitations of the same channel, $ \alpha = 1$, or of another channel, $ \alpha = 2$.

The following symmetry of the fermionic correlation function, $ G ^{(1)}_{}\left(  t_{1}; t_{2} \right) = \left[ G ^{(1)}_{} \left(  t_{2} ; t_{1} \right)  \right]^{*}$, guaranties that $J_{ \alpha}$ is real. 
This is a reason why we can omit a sign of the real part, which is, strictly speaking, should be present on the right hand side of Eq.~(\ref{04e}).  

In fact, $J_{ \alpha}$ depends on the (square) of the overlap integral between the wave functions of an injected particle and of a thermal excitation. 
The difference in signs of  ${\cal P}^{corr}_{34}$ and $ {\cal P}^{ab}_{34}$ is due to different physical consequences of such an overlap. 

\paragraph{The antibunching correction to the shot noise.}
When an electron propagating in the waveguide $ \alpha = 1$ reaches the wave splitter and there overlaps with the thermally excited particles from the waveguide $ \alpha=2$, the quantum statistical, fermionic repulsion forces colliding particles to be scattered into different outputs. 
The colliding fermions, so to say, antibunch while colliding at the wave splitter.  
Therefore, in this case,  overlap leads to more regularized scattering of colliding particles, which suppresses the shot noise. 
This is why ${\cal P}^{ab}_{34}$ has the sign opposite to that of ${\cal P}^{p}_{34}$. 

Formally, the antibunching effect is due to the quantum-mechanical exchange of two fermions, which compete for the same trajectory after scattering. 
This is an electron analogue \cite{Bocquillon:2013dp,Dubois:2013ul,Freulon:2015jo,Glattli:2016wl,Glattli:2016tr,Marguerite:2016ur,Marguerite:2016jt} of the well known in optics Hong-Ou-Mandel effect \cite{Hong:1987gm}.  

\paragraph{The correlation correction to the shot noise.}  
When the source attempts to inject an electron into the waveguide, where another fermions, the thermally excited quasi-particles, are already present, then the quantum-statistical, fermionic repulsion forces an injected electron to be in the state, which is different from the states of present particles. 

At zero temperature, this quantum-statistical repulsion results in the state of an injected electron, which is orthogonal to the state of the Fermi sea. 
The overlap integral in this case is zero and $J_{1} = 0$.

At non-zero temperature, the situation is more subtle. 
Since at non-zero temperature the states of particles are mixed states, the  state of an injected particle is not necessarily orthogonal to the states of existing particles. 
The reason is that each of particles can occupy different components of its mixed state, so as not to be at the same time in the same state with some other particle. 
The mixed state in itself provides more room for fermions to avoid each other.  
As a result, the overlap integral is not zero and $J_{1} \ne 0$, at non-zero temperature.

Since the overlap integral is not zero, the injected particle and the already present thermal excitations together effectively enhance occupation of  states. 
Therefore, the occupation of states impinging the wave splitter from the input $ \alpha=1$ is enhanced, which enhances the shot noise (compare:   ${\cal P}^{p}_{34}$ decreases with increasing temperature, since the mixed state of injected particle contains many components but with small occupation). 
This is why the sign of ${\cal P}^{corr}_{34}$ is the same as the sign of ${\cal P}^{p}_{34}$. 

Formally, the correlation effect is due to the interference of quantum-mechanical amplitudes of two scattered fermions, \cite{Buttiker:1990tn,Blanter:2000wi, Samuelsson:2004uv} which do not compete for  the same trajectory after scattering. 
This is an electron analogue \cite{Henny:1999tb,Oliver:1999ws,Oberholzer:2000wx,Neder:2007jl,Bocquillon:2012if,{Dubois:2013ul}} of the well known in optics Hanbury Brown and Twiss effect \cite{HanburyBrown:1956bi}.

\subsection{How to measure separately three contributions to the excess shot noise} 
\label{sec2d}

When the both waveguides have the same temperature, $ \theta_{1} = \theta_{2} \equiv \theta$, the overlaps integrals become the same, $J_{1} = J_{2}$.
As a result ${\cal P}^{corr}_{34} + {\cal P}^{ab}_{34} =0$ and the excess shot noise is caused solely by scattering of particles generated by the source at temperature $ \theta_{1} = \theta$, ${\cal P}^{ex}_{34} = {\cal P}^{p}_{34}$.

If then we lower the temperature of the second waveguide to zero, $ \theta_{2}= 0$, we eliminate the antibunching correction, ${\cal P}^{ab}_{34} =0$. 
I this case the excess noise is given by the sum of two terms, ${\cal P}^{ex}_{34} = {\cal P}^{p}_{34} + {\cal P}^{corr}_{34}$. 
By subtracting already known ${\cal P}^{p}_{34}$ at temperature $ \theta$ we are left with the correlation correction ${\cal P}^{corr}_{34}$ at temperature $ \theta$.
And finally, as we already mentioned, the correction to the shot noise caused by  antibunching of electrons injected at temperature $ \theta$ with the thermal excitations at temperature  $ \theta$ is opposite in sign to the correlation correction, ${\cal P}^{ab}_{34} = - {\cal P}^{corr}_{34} $. 
Thus, we determined all three contributions. 

To measure  the antibunching correction, rather than a correlation one, we need to start from the setup with both waveguides kept at zero temperature, $ \theta_{1} = \theta_{2} = 0$. 
In this case the excess shot noise provides us with ${\cal P}^{p}_{34}$ at zero temperature, which is  ${\cal P}^{p}_{34}= - {\cal P}^{}_{0}$ in the case when   an electron source emits one particle per period.

Then we increase the temperature of the second waveguide to some finite value, $ \theta_{2} \ne 0$.  
Remind, that a single-electron source injects particles into the first waveguide with zero temperature, $ \theta_{1}=0$, and, therefore, there is no correlation correction to the shot noise, ${\cal P}^{corr}_{34} =0$. 
As a result, the excess shot noise is given by , ${\cal P}^{ex}_{34} = {\cal P}^{p}_{34} + {\cal P}^{ab}_{34}$.  
Provided that ${\cal P}^{p}_{34}$ at zero temperature was already measured, we arrive at ${\cal P}^{ab}_{34}$, which is due to antibunching of particles injected at zero temperature and thermal excitations at some  temperature $ \theta_{2}$. 

To illustrate the general reasoning given above, I consider, as an example, the source of singly-charged levitons emitted at a rate of one particle per period. \cite{{Dubois:2013ul},Dubois:2013fs} 
In the same way, another single-electron source, the quantum capacitor \cite{Feve:2007jx}, can be analyzed. 
The corresponding fermionic correlation function $ G ^{(1)}_{}$ at non-zero temperature was calculated in Ref.~\onlinecite{Moskalets:2017vk}.

\section{Source of levitons}
\label{sec3}

\subsection{Relevant first-order fermionic correlation functions}

The first-order correlation function $ G ^{(1)}_{}$ of excitations in the Fermi sea is defined as the difference of the correlation functions of the Fermi sea with and without these excitations. \cite{{Grenier:2011dv},{Haack:2013ch}}

\subsubsection{Leviton}

The correlation function of a leviton created at a temperature $ \theta$ can be written as follows, \cite{Moskalets:2015ub,Moskalets:2017vk} 

\begin{subequations}
\label{05}
\begin{eqnarray}
G^{(1)}_{L,\theta} (t_{1};t_{2}) &=& 
\int _{- \infty}^{ \infty } d \epsilon \, p_{\theta}( \epsilon) G^{(1)}_{L, \epsilon} (t_{1};t_{2}). 
\label{05a} 
\end{eqnarray}
\ \\ \noindent
Here the derivative of the Fermi distribution function,
\begin{eqnarray}
p_{\theta}( \epsilon) &=& - \frac{ \partial f_{}}{ \partial \epsilon }  =  \frac{ 1 }{ 4 k_{B} \theta \cosh^{2}\left(  \frac{ \epsilon }{ 2 k_{B} \theta } \right)} ,
\label{05c} 
\end{eqnarray}
\ \\ \noindent
defines the probability density for the components of mixed state.
Each such a component is described by the following single-particle correlation function,  
\begin{eqnarray}
G^{(1)}_{ L,\epsilon} (t_{1};t_{2}) &=& 
\frac{ 1 }{ v _{ \mu} }   \Psi_{L, \epsilon}^{*}\left(  t_{1} \right)  \Psi_{L, \epsilon}\left(  t_{2}\right) ,
\nonumber \\
\label{05b} \\
 \Psi_{L, \epsilon}\left(  t \right) &=& e ^{ - i  t \frac{ \mu + \epsilon  }{ \hbar } }  \sqrt{\frac{ \Gamma _{\tau} }{ \pi  }} \frac{ 1 }{ t - i \Gamma _{\tau}  } ,
\nonumber 
\end{eqnarray}
\end{subequations}
with $ \Psi_{L, \epsilon}$ being the wave function of leviton \cite{{Keeling:2006hq},{Dubois:2013fs}} of duration  $ 2\Gamma _{\tau}$ created on top of the (fictitious) Fermi sea with chemical potential $ \mu + \epsilon$ (and with zero temperature). 
This (fictitious) Fermi sea is nothing but a component of the mixed state of the Fermi sea at non-zero temperature, see Eq.~(\ref{Fsea}).  

At zero temperature the probability density is the Dirac delta function, $p_{\theta=0}( \epsilon) = \delta\left(  \epsilon \right)$, and a leviton is created with certainty in a pure single-particle state with $ \epsilon=0$:  $G^{(1)}_{L,\theta=0} = G^{(1)}_{ L,\epsilon=0}$.  
With increasing temperature the probability to find a leviton in the states  with $ \epsilon \ne 0$ increases, and hence the mixedness of its state increases.  

\subsubsection{Thermal excitations}

By analogy, the correlation function of thermally excited quasi-electrons and holes at temperature $ \theta$ can be represented as follows, 
\begin{subequations}
\label{06}
\begin{eqnarray}
G^{(1)}_{th, \theta} (t_{1};t_{2}) &=& 
\int _{- \infty}^{ \infty } d \epsilon \, p_{\theta}( \epsilon) G^{(1)}_{ \epsilon}  (t_{1};t_{2}) , 
\label{06a} 
\end{eqnarray}
where the components of multi-particle mixed state are described by the following correlation function, 
\begin{eqnarray}
G^{(1)}_{ \epsilon}  (t_{1};t_{2}) &=& \frac{1 }{ v_{ \mu}  } \int\limits _{ 0}^{   \epsilon } \frac{ d \epsilon^{\prime} }{ \hbar } \Phi_{ \epsilon^{\prime}}^{*} \left(  t_{1} \right) \Phi_{ \epsilon^{\prime}}\left(  t_{2} \right) ,
\nonumber \\
\label{06b} \\
\Phi_{ \epsilon}(t) &=& \frac{ 1 }{ \sqrt{2 \pi}  } e^{- i t \frac{ \mu + \epsilon }{ \hbar  }} .
\nonumber 
\end{eqnarray}
\end{subequations}
The correlation function $G^{(1)}_{ \epsilon}$ corresponds to a pure multi-particle state, the electron-like state for $ \epsilon > 0$ and the hole-like state for $ \epsilon < 0$. 
This is the same state as the one injected into a zero temperature electron waveguide  when a constant bias $eV = \epsilon$ is applied to its metallic contact. 
This state consists of plane waves $ \Phi_{ \epsilon}$ with energy ranging from $ \mu$ to $ \mu + \epsilon$.  
Since the probability density is an even function of energy, $ p_{ \theta}( \epsilon) = p_{ \theta}( -\epsilon) $, the occupation probabilities of corresponding electron-like and hole-like states, $G^{(1)}_{ \epsilon}$ and $G^{(1)}_{ - \left | \epsilon \right |}$,  are the same.

Note that the correlation function $G^{(1)}_{th, \theta}$ is defined as the difference of Fermi sea correlations functions at temperature $ \theta$  
and at zero temperature, $G^{(1)}_{th, \theta} = G^{(1)}_{F, \theta} - G^{(1)}_{F, \theta=0}$, where  
\begin{eqnarray}
G^{(1)}_{F, \theta} &=& \frac{e ^{ i \left(  t_{1} - t_{2} \right) \frac{ \mu  }{ \hbar } }  }{ 2 \pi i  v_{ \mu}} 
\frac{ 1/  \tau_{ \theta }}{ \sinh\left( \left[   t_{1} - t_{2}  \right]/ \tau_{ \theta}  \right) } 
\nonumber \\
\label{Fsea} \\
&=&
\int _{- \infty}^{ \infty } d \epsilon \, p_{\theta}( \epsilon) 
\frac{1 }{ v_{ \mu} } \int\limits _{ - \infty}^{ \mu + \epsilon } \frac{ d \epsilon^{\prime} }{ \hbar } \Phi_{ \epsilon^{\prime}}^{*} \left(  t_{1} \right) \Phi_{ \epsilon^{\prime}}\left(  t_{2} \right) ,
\nonumber 
\end{eqnarray}
with  the thermal coherence time $ \tau_{ \theta} = \hbar /( \pi k_{B} \theta)$, see, e.g., Ref.~\onlinecite{Grenier:2011js}. 

The representation for $G^{(1)}_{th, \theta}$ given in Eq.~(\ref{06}) is not unique. 
Integration by parts in energy $ \epsilon$ gives a more familiar representation with the Fermi distribution function $f\left(  \epsilon \right)$ (rather than its energy derivative) as the probability density and single-particle plane wave states $\Phi_{ \epsilon}$ (rather then their bunch) as  components of the mixed single-particle state, see, e.g., Refs.~\onlinecite{{Beenakker:2005vx},{Samuelsson:2009voa}}. 
Notice, for holes ($ \epsilon < 0$) the distribution function is $1 - f\left(  \epsilon \right)$. 

In fact, these two representations appeal to different pictures, multi-particle and single-particle, respectively.
For the purposes of this study, the representation given in Eq.~(\ref{06}) is more convenient. 

The same is true for the Fermi sea correlation function $G^{(1)}_{F, \theta}$, Eq.~(\ref{Fsea}).

\subsection{Shot noise caused by scattering of levitons}

Substituting $G ^{(1)}_{p, \theta} = G ^{(1)}_{L, \theta}$, see Eq.~(\ref{05}), into Eq.~(\ref{04b})  we find, \cite{Moskalets:2015ub}
\begin{subequations}
\label{07}
\begin{eqnarray}
\frac{  {\cal P}^{p}_{34} }{  {\cal P}_{0} } &=& - 
\int\limits _{ - \infty }^{ \infty } d \epsilon  p_{\theta}( \epsilon) 
\int\limits _{ - \infty }^{ \infty } d \epsilon^{\prime} p_{\theta}( \epsilon^{\prime})  
{\cal J}\left(  \epsilon, \epsilon^{\prime} \right)  ,
\label{07a} 
\end{eqnarray}
where ${\cal J}\left(  \epsilon, \epsilon^{\prime} \right)$ is given by the square of the overlap integral of the components of the  mixed state of a leviton with energy $ \epsilon$ and $ \epsilon ^{\prime}$,
\begin{eqnarray}
{\cal J}\left(  \epsilon, \epsilon^{\prime} \right)   &=& \left | \int _{}^{ } dt_{} 
\Psi_{L, \epsilon}^{*}\left(  t_{} \right)
\Psi_{L, \epsilon^{\prime}}^{}\left(  t_{} \right)
 \right |^{2} =
e^{- \frac{ \left | \epsilon - \epsilon^{\prime} \right | }{  {\cal E}_{L}  }} , 
\label{07b} 
\end{eqnarray}
with ${\cal E}_{L} = \hbar/ (2 \Gamma _{\tau})$ being the energy of a leviton. \cite{Keeling:2006hq}

The temperature dependence of ${\cal P}_{34}^{p}$, Eq.~(\ref{07a}), is shown in Fig.~\ref{fig2}. 
\begin{figure}[b]
\includegraphics[width=80mm, angle=0]{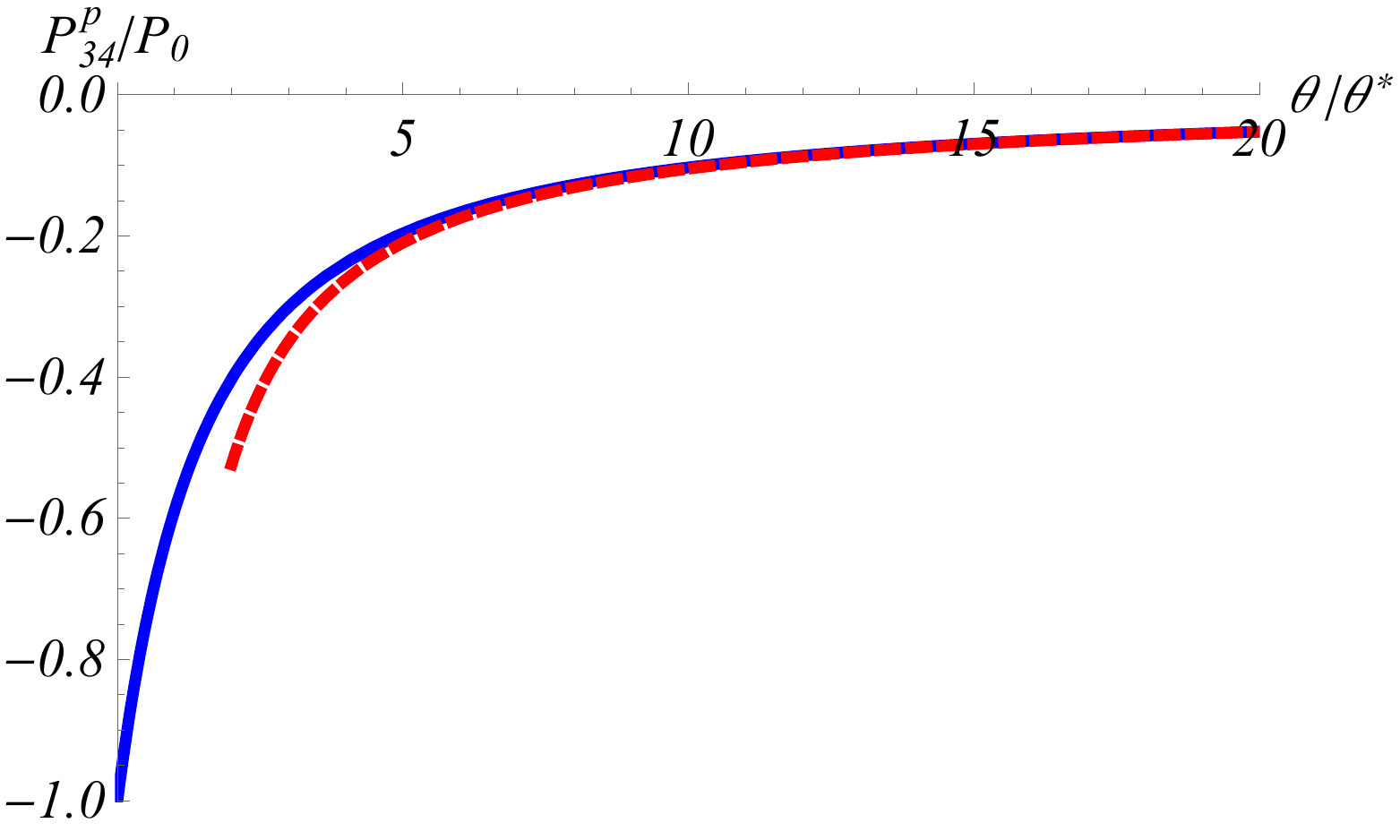}
\caption{The shot noise, ${\cal P}_{34}^{p}$, Eq.~(\ref{07a}), (a blue solid line), caused by partitioning a leviton per period ${\cal T}$ on a wave splitter with transmission probability $T$, is shown as a function of the  temperature $ \theta$, at which a leviton is created. 
The constant $ {\cal P}_{0} = e^{2}T\left(  1-T \right)/ {\cal T}$ is a (minus) zero-temperature shot noise. 
The characteristic temperature $k_{B} \theta^{\star} = {\cal E}_{L}/\pi$ is defined by the energy ${\cal E}_{L}= \hbar /(2 \Gamma _{\tau})$ of a leviton of duration $ 2 \Gamma _{\tau}$. 
The high temperature asymptotics, Eq.~(\ref{07c}), is shown as a red dashed line.
}
\label{fig2}
\end{figure}
For low temperatures, $ \theta \ll \theta^{*} $ with $k_{B} \theta^{\star} = {\cal E}_{L}/\pi$ (or, equivalently, for long thermal coherence times,  $ \tau_{ \theta} \gg 2\Gamma _{\tau}$), when one can use $p\left(   \epsilon \right) \approx \delta\left(  \epsilon \right)$, the noise tends to the shot noise of a single particle per period, $ {\cal P}^{p}_{34}   = - {\cal P}_{0}$.

While at high temperatures it gets suppressed.  
To estimate the shot noise one can proceed as follows. 
At $ \theta \gg \theta^{*}$, when $p_{\theta}\left(  \epsilon \right)$ and $p_{\theta}\left(  \epsilon^{\prime} \right)$ in Eq.~(\ref{07a}) are almost constant over the energy interval of order $ {\cal E}_{L}$, the overlap integral acts like a delta function, ${\cal J}\left(  \epsilon, \epsilon^{\prime} \right) \approx 2 {\cal E}_{L} \delta\left( \epsilon - \epsilon^{\prime} \right)$.  
This means, that the shot noise can be represented as the sum of contributions due to different components of the mixed state, which are correlated within an energy interval $2 {\cal E}_{L}$, but not correlated otherwise: \cite{Moskalets:2015ub}

\begin{eqnarray}
\left.
\frac{  {\cal P}^{p}_{34} }{  {\cal P}_{0} } \right |_{ \theta \gg \theta^{*}}
&\approx& - 2 {\cal E}_{L} 
\int\limits _{- \infty}^{ \infty} d \epsilon p_{\theta}^{2}( \epsilon) = - \frac{ \pi }{ 3  } \frac{ \theta^{*} }{ \theta  } .
\label{07c} 
\end{eqnarray}
\end{subequations}
\ \\ \noindent
This asymptotics is shown as a red dashed line in Fig.~\ref{fig2}. 
Remind that $p_{ \theta}( \epsilon)$ is the probability (density) of appearance of the component of mixed state with energy $ \epsilon$. 
The square of this probability determines the contribution of this component to  single-particle shot noise. \cite{Moskalets:2011jx,Moskalets:2015ub}   

Thus, the gradual suppression of ${\cal P}^{p}_{34}$ with increasing temperature indicates increasing mixedness of a single-particle quantum state.

\subsection{Correlation correction to the shot noise}

Using Eqs.~(\ref{05}) and (\ref{06}) in Eqs.~(\ref{04c}) and ~(\ref{04e})  we calculate the correlation correction,  

\begin{subequations}
\label{08}
\begin{eqnarray}
\frac{  {\cal P}^{corr}_{34} }{  {\cal P}_{0} } &=& - 2
\int\limits _{ - \infty}^{ \infty } d \epsilon  p_{\theta}( \epsilon) 
\int\limits _{ - \infty}^{ \infty } d \epsilon^{\prime} p_{\theta}( \epsilon^{\prime})  
\int\limits _{0}^{ \epsilon^{\prime} } \frac{ d \epsilon^{\prime\prime}   }{ \hbar  }
f_{L, \epsilon}( \epsilon^{\prime\prime}) ,
\nonumber \\
\label{08a} \\
f_{L, \epsilon}( \epsilon^{\prime\prime})   &=&
\left | \int _{}^{ } dt_{} 
\Psi_{L, \epsilon}^{}\left(  t_{} \right)
\Phi_ {\epsilon^{\prime\prime} }^{*}\left(  t_{} \right) \right |^{2} 
\nonumber \\
&=& \left | \int _{}^{ } \frac{ dt_{}  }{ \sqrt{2 \pi}  }
\sqrt{\frac{ \Gamma _{\tau} }{ \pi  }} \frac{ 1 }{ t - i \Gamma _{\tau}  }
e^{ i t \frac{ \epsilon^{\prime\prime} -  \epsilon  }{ \hbar  }} \right |^{2} 
\nonumber \\
&=& 
\theta\left(  \epsilon^{\prime\prime } - \epsilon \right) 2 \Gamma _{\tau} 
e^{- \frac{ \left ( \epsilon^{\prime\prime} - \epsilon \right ) }{  {\cal E}_{L}  }} .
\label{08b} 
\end{eqnarray}
\noindent \\
Here $ \theta(x)$ is the Heaviside step function. 
The quantity $f_{L, \epsilon}( \epsilon^{\prime\prime})$, which is given by the square of the Fourier coefficient of the wave function of a leviton $ \Psi_{L, \epsilon}$, has the following meaning. 
Mathematically, this is nothing but the energy distribution function of a leviton \cite{Keeling:2006hq,Moskalets:2014ea} created on top of the  Fermi sea with zero temperature and the Fermi energy  $ \mu + \epsilon$. 
However, within the physical context of the problem in study, this quantity characterizes the overlap between a leviton and the thermal excitations of the Fermi sea.  
To be more precise,  the integral over $ \epsilon ^{\prime\prime}$ is the overlap between the component $ \Psi_{L, \epsilon}$ of the mixed state of a single leviton and the component, described by $G^{(1)}_{ \epsilon}$, of the multi-particle mixed state of the thermal excitations of the Fermi sea.

The peculiar property of a leviton  is that the distribution function $f_{L, \epsilon}( \epsilon^{\prime\prime})$ is non-zero only for energies $E = \mu + \epsilon^{\prime\prime}$ larger than the corresponding Fermi energy $ \mu + \epsilon$, that is, for $ \epsilon^{\prime\prime} > \epsilon$. 
This is reflected by the Heaviside step function in Eq.~(\ref{08b}).    
The distribution function is normalized such that $ \int _{ \epsilon}^{ \infty } (d \epsilon^{\prime\prime}/ \hbar) f_{L, \epsilon}( \epsilon^{\prime\prime}) = 1$. 

The temperature dependence of ${\cal P}_{34}^{corr}$, Eq.~(\ref{08}), is shown in Fig.~\ref{fig3}. 
First of all, the sign of the correction ${\cal P}_{34}^{corr}$ is the same as the one of the main contribution, ${\cal P}_{34}^{p}$, Eq.~(\ref{07a}). 
Therefore, the correlations that arise, when a leviton is created, do enhance the measured noise.  
The non-monotonous behavior of ${\cal P}_{34}^{corr}$ as a function of temperature is explained as follows.

\begin{figure}[b]
\includegraphics[width=80mm, angle=0]{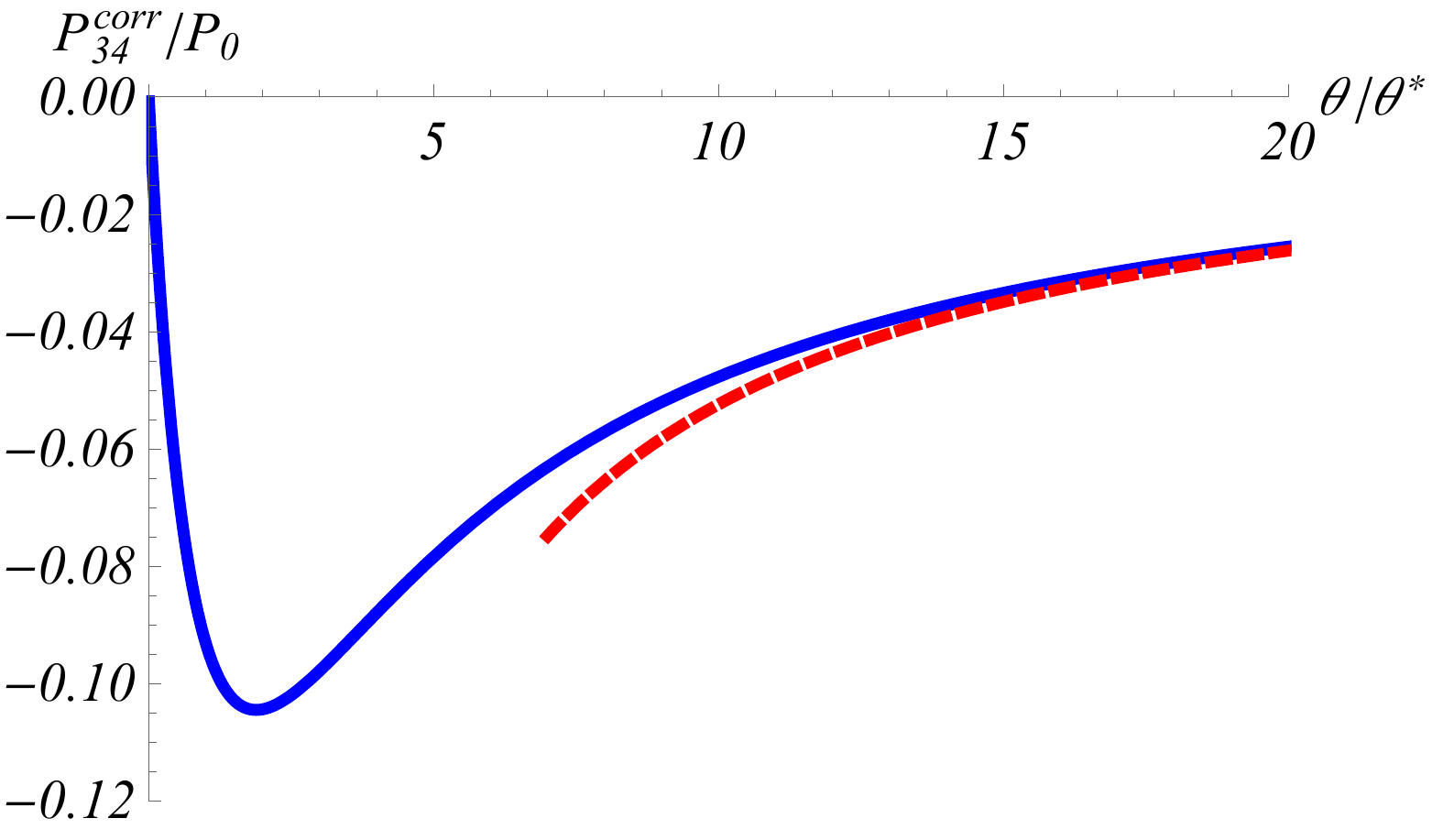}
\caption{The correlation correction to the shot noise of levitons, ${\cal P}^{corr}_{34}$, Eq.~(\ref{08}), (a blue solid line) is shown as a function of temperature $ \theta$. 
The high temperature asymptotics, Eq.~(\ref{08c}), is shown as a red dashed line.
The units are the same as in Fig.~\ref{fig2}. 
}
\label{fig3}
\end{figure}

For low temperatures,  $ \theta \to 0$, there are no thermal excitations and there is nothing to be correlated with. 
The correlation correction is zero. 
At intermediate temperatures, $ \theta \sim \theta^{*}$ , the effective energy of all the thermal excitations is of the order of the leviton's energy, $ k_{B} \theta^{} \sim {\cal E}_{L}/ \pi $, and the correlations are maximum.  
The suppression of ${\cal P}_{34}^{corr}$ at high temperatures,  $ \theta \gg \theta^{*}$, has the same origin as the suppression of ${\cal P}_{34}^{p}$, that is, it is caused by the increasing mixedness of the quantum state of leviton. 

Interestingly, the asymptotics of ${\cal P}_{34}^{corr}$ at high temperatures is exactly half of that of ${\cal P}_{34}^{p}$, Eq.~(\ref{07c}), 

\begin{eqnarray}
\left.
\frac{  {\cal P}^{corr}_{34} }{  {\cal P}_{0} } \right |_{ \theta \gg \theta^{*}}
&\approx&  - \frac{ \pi }{ 6  } \frac{ \theta^{*} }{ \theta  } .
\label{08c} 
\end{eqnarray}
\end{subequations}
\ \\ \noindent
This dependence if shown in Fig.~\ref{fig3} as a red dashed line. 

Such a halving is explained as follows. 
At high temperatures, $ k_{B}\theta \gg {\cal E}_{L}$, there are a lot of thermal excitations with various energies, such that the strength of the correlations arising is restricted only by the capability of a leviton. 
This is why at high temperatures the exponential factors of both quantities, $f_{L, \epsilon}( \epsilon^{\prime\prime}) $ in Eq.~(\ref{08}) and ${\cal J}\left(  \epsilon, \epsilon^{\prime} \right)$ in Eq.~(\ref{07}), can be approximated by the same delta function. 
As a consequence, the correlation correction ${\cal P}_{34}^{corr}$, Eq.~(\ref{08a}), is  related to ${\cal P}_{34}^{p}$, Eq.~(\ref{07a}). 

The factor one half comes from the fact that a leviton with wave function $\Psi_{L, \epsilon}^{}\left(  t_{} \right)$ is correlated only with those thermal excitations, whose energy exceeds $ \mu + \epsilon$, see Eq.~(\ref{08b}). 
A leviton is not correlated with those thermal excitations, whose energy is less than the chemical potential $ \mu + \epsilon$.   
For brevity, one can call them (effective) quasi-electrons and holes, respectively. 
Quasi-electrons and holes appear on the average with an equal probability and, therefore, only in half the cases a levitons is subject to correlations. 
This reasoning explains why the reduction factor is exactly $1/2$. 

Note that another example of particles with the distribution function $f_{L, \epsilon}( \epsilon^{\prime\prime})$, Eq.~(\ref{08b}), are electrons injected by a quantum capacitor\cite{Feve:2007jx} in the appropriate regime\cite{Keeling:2008ft,Moskalets:2017vk}. 
For this source, we can expect an analogous temperature dependence of the correlation noise.

\subsection{Antibunching correction to the shot noise}
\label{sec3d}

\begin{figure}[b]
\includegraphics[width=80mm, angle=0]{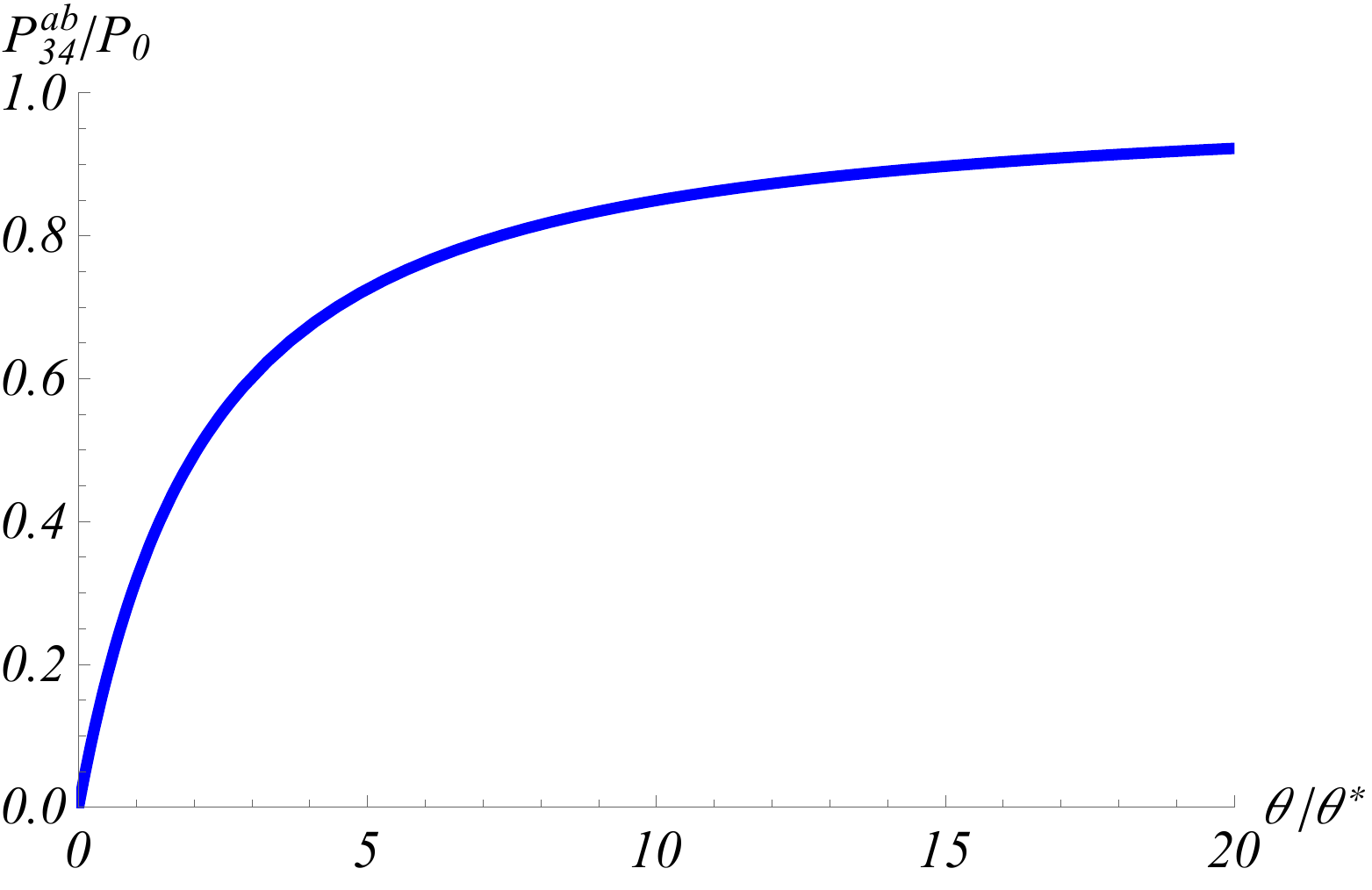}
\caption{The antibunching correction to the shot noise of a leviton created at zero temperature, ${\cal P}^{ab}_{34}$, Eq.~(\ref{09b}), is shown as a function  of temperature of thermal excitations, $ \theta_{2} =  \theta$. 
The units are the same as in Fig.~\ref{fig2}.
}
\label{fig4}
\end{figure}

The antibunching correction ${\cal P}^{ab}_{34}$, Eq.~(\ref{04d}), is determined by $J_{2}$, Eq.~(\ref{04e}), which generally depends on two temperatures, $ \theta_{1}$,  the temperature at which a leviton is created, and $ \theta_{2}$,  the temperature of thermal excitations, which a leviton collides on a wave splitter with. 
Therefore, while calculating $J_{2}$ we use Eq.~(\ref{05}) with $ \theta = \theta_{1}$ and Eq.~(\ref{06}) with $ \theta = \theta_{2}$. 
Then the antibunching correction to the shot noise of levitons, ${\cal P}^{ab}_{34}$, Eq.~(\ref{04d}), becomes 
\begin{subequations}
\label{09}
\begin{eqnarray}
\frac{  {\cal P}^{ab}_{34} }{  {\cal P}_{0} } &=&  2
\int\limits _{ - \infty}^{ \infty } d \epsilon  p_{\theta_{1}}( \epsilon) 
\int\limits _{ - \infty}^{ \infty } d \epsilon^{\prime} p_{\theta_{2}}( \epsilon^{\prime})  
\int\limits _{0}^{ \epsilon^{\prime} } \frac{ d \epsilon^{\prime\prime}   }{ \hbar  }
f_{L, \epsilon}( \epsilon^{\prime\prime})  ,
\nonumber \\
\label{09a}
\end{eqnarray}
with the distribution function of levitons, $f_{L, \epsilon}( \epsilon^{\prime\prime})$, defined in Eq.~(\ref{08b}). 
Notice the different signs in Eqs.~(\ref{07a}) and (\ref{09a}). 
Therefore, antibunching suppresses the shot noise. \cite{{Liu:1998wr},Bocquillon:2013dp,Dubois:2013ul}

The antibunching correction differs from the correlation corrections also in   another reason.
The correlation correction, ${\cal P}^{corr}_{34}$, Eq.~(\ref{08a}), vanishes when a leviton is created at zero temperature, $ \theta_{1}=0$. 
While, the antibunching correction,  ${\cal P}^{ab}_{34}$, Eq.~(\ref{09a}),  survives the limit $ \theta_{1} \to 0$, as long as the temperature of the other contact is not zero, $ \theta_{2} > 0$.  

Indeed, when $ \theta_{1}=0$ we use $ p_{\theta_{1}=0}( \epsilon) = \delta\left(  \epsilon \right)$ in Eq.~(\ref{09a}) and obtain, 

\begin{eqnarray}
\frac{  {\cal P}^{ab}_{34} }{  {\cal P}_{0} } &=&
1 - 2 \int _{0}^{ \infty } d \epsilon^{\prime} \, p_{\theta, 2}( \epsilon^{\prime})  e^{ - \frac{ \epsilon^{\prime} }{ {\cal E}_{L}  } } .
\label{09b} 
\end{eqnarray}
\end{subequations}
\noindent \\
The temperature dependence of this correction is shown in Fig.~\ref{fig4}. 
Remind that in this case the shot nose caused by a leviton  is ${\cal P}^{p}_{34} = - {\cal P}^{}_{0}$, see Eq.~(\ref{07a}) at $ \theta=0$. 

At low temperatures, $ \theta_{2} \to 0$, there are no thermally excited quasi-particles capable to antibunch with a leviton and, therefore, ${\cal P}^{ab}_{34} \to 0$. 
While with increasing temperature, $ \theta_{2} \geq \theta^{*}$ the antibunching noise suppression becomes essential. 
The maximum suppression occurring when $ \theta_{2} \gg \theta^{*}$  is as large as ${\cal P}^{}_{0}$. 

This suppression is interpreted as follows. 
When the two colliding electrons overlap perfectly on the wave splitter, then both them do not contribute to the shot noise. 
This is valid for periodic \cite{Olkhovskaya:2008en,Moskalets:2011jx,Splettstoesser:2009im,Splettstoesser:2010bo,Jonckheere:2012cu,Dubois:2013fs,Bocquillon:2013fp,Ferraro:2014if,Ferraro:2014ux,Vanevic:2016eq,Rech:2016cd,Hofer:2017hj,Rech:2017be} as well as random \cite{Moskalets:2011jx,Glattli:2017vp} single-electron injection.  
Therefore, in the case of two colliding levitons, the total shot noise is   suppressed by $2 {\cal P}^{}_{0}$.\cite{Dubois:2013ul} 
If the colliding particles overlap only partially, then their contribution to the shot noise is suppressed only partially.  
In any case, only $1/2$ of the antibunching suppression is attributed to the  suppression of the contribution of one of the colliding particles. 

Therefore, in our case, since the total suppression is ${\cal P}^{}_{0}$, only half of the contribution of a leviton to the shot noise is suppressed. 
The other half of ${\cal P}^{}_{0}$ is attributed to the suppression of the contribution of thermal excitations.  

Why the leviton's contribution is not fully suppressed in the high temperature limit, when there are a lot of thermally excited quasi-particles? 
The answer is following.
A leviton created at zero temperature is able to overlap only with quasi-electron thermal excitations, that is, with particles describing by the correlation function $G^{(1)}_{ \epsilon}$, Eq.~(\ref{06b}), for $ \epsilon > 0$. 
The quasi-electron excitations appears with probability $1/2$, which is given by $ \int _{0}^{ \infty } d \epsilon p_{ \theta}( \epsilon) = 1/2$, see Eq.~(\ref{06a}). 
The other half is a probability for a hole state to appear, which does not antibunch with a leviton. 
Therefore, a leviton experiences antibunching in half instances only. 
Consequently, its contribution to shot noise is suppressed at best only half.

\section{Discussion and conclusion}
\label{concl}

The quantum state of a single electron created on top of the Fermi sea is affected significantly by temperature. 
I discussed here the consequences of this for electrical shot noise. 
Three effects of temperature on shot noise were identified. 
The first effect is related to modification of a single-particle contribution to shot noise. 
While two other effects are associated with the appearance of two-particle contributions to shot noise.

{\it First effect}: The quantum state of a particle injected at a non-zero temperature  is a mixed state. 
The different components of the mixed state represents alternative (but not orthogonal) histories of a particle. 
Roughly speaking, the contribution to the shot noise of each component  is proportional to the square of the probability of occurrence  of this component. 
Although the sum of all probabilities is one, the sum of the squares of probabilities is less than one.  
As a result the single-particle shot noise gets suppressed.  
A decrease in the shot noise with increasing temperature indicates increasing mixedness of a single-particle quantum state.

{\it Second effect}: The state of a particle injected at a non-zero temperature is not orthogonal to the state of the Fermi sea. 
As a result, the thermal excitations and a particle  injected by a  single-electron source demonstrate Hanbury Brown and Twiss correlations that enhance shot noise. 
The temperature behavior of this contribution is non-monotonic. 

At low temperatures, the increase in this contribution is caused by an increase in the number of thermal excitations. 
While at high temperatures, the corresponding contribution decreases with temperature. 
Using the source of levitons as an example, I demonstrated that this decrease is exactly half the same as a decrease of the single-particle shot noise. 
The reduction factor comes from the fact that particles injected on top of the Fermi sea are correlated with quasi-electrons and they do not correlate with holes. 
The quasi-electrons and holes appear with the same probability, hence, the reduction factor is one half. 

{\it Third effect}: The antibunching of particles injected by a single-electron source in the waveguide $ \alpha = 1$ with possibly present thermally excited particles in  the waveguide $ \alpha=2$, does suppress the measured noise. 
This suppression is an electron analogue of the optical Hong-Ou-Mandel effect.
The contribution of this third effect to noise is described by the second-order coherence analogous to that, which describes the second effect, but with the opposite sign. 
Consequently, in the case of a single-particle colliding with thermal excitations, the strength of Hong-Ou-Mandel correlations is maximally limited to half of what a single-particle would demonstrate if it would collide with an identical particle instead. 

The difference from the second effect, however, is that now the second-order coherence  depends on two temperatures: the temperature $ \theta_{1}$, which the injected particle was created at, and the temperature $ \theta_{2}$ of thermal excitations.     
As a result, the antibunching effect exists, even if the injected particle was created  in a pure state at zero temperature, $ \theta_{1}=0$, provided that $ \theta_{2} \ne 0$. 

In conclusion, I analyzed in detail the temperature dependence of the shot noise caused by particles injected onto the surface of the Fermi sea by a single-electron source. 
Such a dependence is ultimately due to the fact that the mixedness of quantum state of injected particles increases with increasing temperature. 
An experimental proof of the predicted effects would demonstrate that the shot noise is effective for experimental study of mixed quantum states.

\acknowledgments

I thank Gwendal F\`{e}ve for helpful and stimulating discussions. 
I appreciate the warm hospitality of the Aalto University, Finland, where part of this work was accomplished. 


\end{document}